\documentclass{JHEP3} 
\def\be{\begin{equation}}
\def\ee{\end{equation}}
\def\bc{\begin{center}}
\def\ec{\end{center}}
\def\bea{\begin{eqnarray}}
\def\eea{\end{eqnarray}}
\def\dd{\displaystyle}
\def\nn{\nonumber}

\def\ad{\dot{\alpha}}
\def\ov{\overline}
\def\cP{{P}}
\title{Brane-induced supersymmetry breaking}
\author{Jonathan Bagger \\ Department of Physics and 
Astronomy, The Johns Hopkins University, \\
3400 North Charles Street, Baltimore, MD 21218, USA
\\ E-mail: \email{bagger@jhu.edu}}
\author{Ferruccio Feruglio \\
Dipartimento di Fisica `G.~Galilei', Universit\`a di Padova and
\\
INFN, Sezione di Padova, Via Marzolo~8, I-35131 Padua, Italy
\\ 
E-mail: \email{feruglio@pd.infn.it}}
\author{Fabio Zwirner \\
Dipartimento di Fisica, Universit\`a di Roma `La Sapienza' and
\\
INFN, Sezione di Roma, P.le Aldo Moro~2, I-00185 Rome, Italy
\\
E-mail: \email{fabio.zwirner@roma1.infn.it}}
\preprint{\hepth{0108010v2}}
\abstract{
We study spontaneous supersymmetry breaking induced by brane-localized
dynamics in five-dimensional supergravity compactified on $S^1/Z_2$.
We consider a model with gravity in the bulk and matter localized
on tensionless branes at the orbifold fixed points.  We assume that
the brane dynamics give rise to effective brane superpotentials that trigger
the supersymmetry breaking.  We analyze in detail the super-Higgs effect.
We compute the full spectrum and show that the symmetry breaking is
spontaneous but nonlocal in the fifth dimension.  We demonstrate that
the model can be interpreted as a new, non-trivial implementation of
a coordinate-dependent Scherk-Schwarz compactification.
}
\begin{document} 
\section{Introduction}

Models with extra dimensions have attracted considerable attention 
because they provide a geometrical approach to the hierarchy 
problems that afflict modern particle physics: the gauge hierarchy 
and the cosmological constant.  It is widely believed that (broken) 
supersymmetry may play a role in generating and stabilizing these 
hierarchies.

In this paper we study supersymmetry breaking induced by brane-localized
dynamics in higher dimensional theories \cite{hor,mno}. (Related work
can be found in \cite{bulk,aqgau}).  For simplicity we consider
compactifications from five to four dimensions on an $S^1/Z_2$ orbifold,
but the mechanism we describe can be extended to higher dimensions.
We assume that the bulk contains pure five dimensional supergravity;
the discussion can be generalized to include bulk vector multiplets
and hypermultiplets as well.  We imagine that tensionless branes
are placed at the orbifold fixed points.  The branes do not generate
a warp factor, so the bosonic vacuum is flat.

We start by writing down a five-dimensional action that describes
bulk supergravity interacting with the vevs of the brane superpotentials.
The action is invariant under the full set of supersymmetries that are
consistent with the orbifold construction.  Since the fifth dimension is
compactified on an orbifold, the five-dimensional supersymmetries
split into an infinite number of four dimensional supersymmetries.  All
but one are nonlinearly realized.

We then study how brane-localized matter can spontaneously break the
remaining $N=1$ supersymmetry.  Our construction is independent of the
details of the brane physics; we simply assert that each brane has an
effective superpotential, the remnant of some localized brane
dynamics.  We assume that each superpotential receives a constant
expectation value (vev).  The superpotential vevs do not affect the
brane tensions, so our construction describes a scenario with
vanishing $F$- and $D$-terms for the brane-localized matter.  The
construction reproduces the main features of gaugino condensation
\cite{hor} in $M$-theory \cite{hw}, at the level of an effective
five-dimensional Lagrangian.

The initial part of our discussion follows, albeit in a simpler context,
the treatment of ref.~\cite{mno}.  However,
we go further in two important respects.  First, we provide a complete
analysis of the super-Higgs effect and derive the spectrum for all
the Kaluza-Klein modes.  Second, we apply the results of \cite{bfz}
and demonstrate that the model can be interpreted as a new type
of coordinate-dependent Scherk-Schwarz compactification
\cite{ss}.  We show that our brane action induces a set of generalized
boundary conditions on the gravitino fields, with field discontinuities
at the orbifold fixed points.

At low energies, our five-dimensional theory reduces to 
spontaneously broken no-scale supergravity \cite{noscale}, 
provided the supersymmetry-breaking mass splittings are small 
relative to the Kaluza-Klein scale.  The order parameter for
supersymmetry breaking is nonlocal in the fifth
dimension \cite{hor}, and is proportional to the average of 
the superpotential vevs on the branes.  We illustrate how this
non-locality ameliorates the instability problems \cite{insta} 
of standard no-scale supergravities~\cite{noscale}.
(This was previously stressed in~\cite{uvsoft} for conventional
Scherk-Schwarz compactifications.)

\section{The bulk action and its spectrum}

Our starting point is pure five-dimensional Poincar\'e supergravity
\cite{pure5d} in its on-shell formulation.
The supergravity multiplet contains the f\"unfbein $e_M^{\;\;A}$, the
gravitino $\Psi_M$ and the graviphoton $B_M$. The five-dimensional
bulk Lagrangian reads~\footnote{The physical dimensions of the
fields depend, of course, on the powers of $M_5$ that are inserted
in the different terms of ${\cal L}_{bulk}$.  Our conventions are
chosen so that all fields have canonical dimension after compactification
from five to four dimensions.}
\bea
\kappa {\cal L}_{bulk}
& = & - {1 \over 2 \kappa^2} e_5 R_5  - {1 \over 4} e_5 F_{MN} F^{MN}
- {\kappa \over 6 \sqrt{6}} \epsilon^{MNOPQ} F_{MN} F_{OP} B_Q \nn \\
&   & + i  \epsilon^{MNOPQ} \ov{\Psi}_M \Sigma_{NO} D_P \Psi_Q - i
\sqrt{3 \over 2} {\kappa \over 2} e_5 F_{MN} \ov{\Psi}^M \Psi^N \nn \\
& & + i \sqrt{3 \over 2} {\kappa \over 4} \epsilon^{MNOPQ} F_{MN}
\ov{\Psi}_O \Gamma_P \Psi_Q + {\rm 4 \!\! - \!\! fermion \;\; terms} \, .
\label{lbulk}
\eea
Unless otherwise stated, our five-dimensional notation is identical
to ref.~\cite{baggerrs}; our four-dimensional notation is the same as
in  ref.~\cite{wb}. In particular, the five-dimensional coordinates are
$x^M =(x^m,x^5)$; $\hat{5}$ is the fifth tangent-space index; $M_5
\equiv \kappa^{-1}$ is the (reduced) five-dimensional Planck mass;
$e_5 = \det e_M^{\;\;A}$; $R_5$ is the five-dimensional scalar
curvature; $e_4 = \det e_m^{\;\;a}$, where the latter are the
components of the f\"unfbein with four-dimensional indices;
and $\epsilon^{MNOPQ} = e_5 e_A^{\;\;M} e_B^{\;\;N}
e_C^{\;\;O} e_D^{\;\;P} e_E^{\;\;Q} \epsilon^{ABCDE}$,
$\epsilon^{mnop} = e_4 e_a^{\;\;m} e_b^{\;\;n}
e_c^{\;\;o} e_d^{\;\;p} \epsilon^{abcd}$, $\epsilon^{\hat{0}
\hat{1} \hat{2} \hat{3} \hat{5}} = \epsilon^{\hat{0} \hat{1}
\hat{2} \hat{3}} = + 1$.

It is not hard to check that the bulk supergravity Lagrangian
is invariant, up to a total derivative, under the following
supersymmetry transformations,
\bea
\delta e_M^{\;\;A}
& = &
i \kappa
\left( \ov{\eta} \Gamma^A \Psi_M -
\ov{\Psi}_M \Gamma^A \eta \right) \, ,
\nn \\
\delta B_M
& = &
- i \sqrt{3 \over 2}
\left( \ov{\eta} \Psi_M -
\ov{\Psi}_M \eta \right) \, ,
\nn \\
\delta \Psi_M
& = &
{2 \over \kappa} D_M \eta - \sqrt{2 \over 3}
\left( \Gamma^N F_{NM} - {1 \over 4} e_5
\epsilon_{MNOPQ} F^{NO} \Sigma^{PQ} \right) \eta
\nn \\[2mm] & & + {\rm 3 \!\! - \!\! fermion \;\; terms} \, .
\label{susygr}
\eea
The transformation parameter $\eta(x^M)$ is a five-dimensional Dirac
spinor. Here and hereafter, we neglect all three- and four-fermion
terms.

In what follows, we take the fifth dimension to be compactified
on the orbifold  $S^1/Z_2$, obtained by the identification $x^5
\leftrightarrow - x^5$. For convenience, we choose to work
on the orbifold covering  space, so we let $x^5$ vary in the
interval $[-\pi \kappa ,\pi \kappa]$.  We define the bulk
action to be~\footnote{Note that the
limits of the integration over the fifth coordinate are set by
convention:
If we let $x^5$ vary between $-\pi/m$ and $\pi/m$, the physical
compactification radius is $R=r/m$. We choose to set $m=M_5$.}
\be
S_{bulk} = {1 \over 2} \,
\int d^4 x \int_{- \pi \kappa}^{+ \pi \kappa}
d x^5 \, {\cal L}_{bulk} \, ,
\label{sbulk2}
\ee
where the factor of $(1/2)$ avoids double-counting equivalent
points.

We take our fields to be fluctuations off the following background,
\be
\langle g_{MN} \rangle =
\left(
\begin{array}{cc}
\eta_{mn} & 0
\\
\phantom{bla}
\\
0 & r^2
\end{array}
\right) \, ,
\label{backg}
\ee
where $r \ne 0$ is an undetermined real constant and all
other background fields are assumed to vanish.  (In principle, $\langle
B_5 \rangle$ is a second undetermined real constant, but we
set it to zero as well.)  This background is a
solution to the five-dimensional equations of motion.  From
(\ref{backg}) we deduce the relation between the four-dimensional
Planck mass $M_4$  and $M_5 \equiv \kappa^{-1}$,
\be
M_4^2 = \pi R \cdot M_5^3 \, ,
\ee
where $R = r/M_5$ is the physical compactification radius.

We define the action of the orbifold symmetry in such
a way that the action, the transformation laws and the background
are all invariant.  Decomposing the five-dimensional spinor $\Psi$,
and its conjugate $\ov{\Psi}$, into four-dimensional form, and
following the convention that $\Psi \equiv (\psi^1_\alpha \, , \,
\ov{\psi^2}^{\ad})^T$ and $\ov{\Psi} \equiv ( \psi^{2 \, \alpha}
\, , \, \ov{\psi^1}_{\ad} )$, we assign even $Z_2$-parity to
\be
e_m^{\;\;a} \, ,
\;\;\;
e_{5 \hat{5}} \, ,
\;\;\;
B_5 \, ,
\;\;\;
\psi_m^1 \, ,
\;\;\;
\psi_5^2 \, ,
\;\;\;
\eta^1 \, ,
\label{peven}
\ee
and odd $Z_2$-parity to
\be
e_5^{\;\;a}
\, , \;\;\;
e_{m \hat{5}}
\, , \;\;\;
B_m
\, , \;\;\;
\psi_m^2
\, , \;\;\;
\psi_5^1
\, , \;\;\;
\eta^2 \, .
\label{podd}
\ee

{}From a four-dimensional point of view, the physical spectrum
contains one massless $N=1$ gravitational multiplet, with
spins $(2,3/2)$, built from the zero modes of $e_m^{\;\;a}$ and
$\psi_m^1$; one massless $N=1$ chiral multiplet, with spins $(1/2,0)$,
composed of the zero modes of $\psi_5^2$, $e_{5 \hat{5}}$ and $B_5$;
and an infinite series of (short) massive multiplets of $N=2$
supergravity, with spins $(2,3/2,3/2,1)$ and squared masses
\be
M_n^2 =  {n^2 \over R^2} \, ,
\;\;\;\;\;
(n=1,2,\ldots) \, .
\label{kkmass}
\ee
The Kaluza-Klein tower gains mass through an infinite series of
Higgs and super-Higgs effects, each occurring at its own mass
level \cite{dolan}. The Kaluza-Klein gravitons and graviphotons
gain mass by eating the Fourier modes of the fields $g_{m5}$,
$g_{55}$ and $B_5$, while the massive gravitinos eat the Fourier
modes of the field $\Psi_5$. This is consistent with the fact that
the parameter of five-dimensional supersymmetry, $\eta(x^M)$,
has an infinite number of Fourier modes.  Each of the modes
generates a supersymmetry; in the absence of matter, all but
one are spontaneously broken. The broken supersymmetries
implement an infinite series of super-Higgs effects for the
massive gravitinos.
The remaining supersymmetry is the $N=1$ supersymmetry
of the four-dimensional low-energy effective action.

\section{Brane action and modified supersymmetry transformations}

Having described the bulk action, we now construct the brane
action.  We are not interested in the details of the brane
physics, so we imagine that the brane fields are integrated
out, leaving a constant superpotential vev on each brane.  We
assume the physics is such that the tension vanishes on each
brane.  We shall see that the superpotential vevs can
spontaneously break the remaining $N=1$ supersymmetry.

In general, the physics is different on the two branes,
so the superpotential vevs need not be the same. The 
brane action we adopt, in analogy with \cite{hor,mno}, 
is
\be
\label{sbrane2}
S_{brane} = {\kappa^2 \over 2} \,
\int d^4 x \int_{- \pi \kappa}^{+ \pi \kappa}
d x^5 \,  e_4 \left[ \delta(x^5)
\ov{P_0} + \delta(x^5 - \pi \kappa) \ov{P_{\pi}}  \right]
\, \psi_a^1 \sigma^{ab} \psi_b^1 +
{\rm \; h.c.} \, ,
\ee
where $P_0$ and $P_{\pi}$ are complex constants with the dimension of
$(mass)^3$ which parametrize the vevs of the superpotentials.

It is not hard to compute the variation of the brane action
using the transformations inherited from the bulk.  We find
\bea
\delta S_{brane} & = & {\kappa^2 \over 2} \,
\int d^4 x \int_{- \pi \kappa}^{+ \pi \kappa}
d x^5 \, e_4  \left[ \delta(x^5)
\ov{P_0} + \delta(x^5 - \pi \kappa) \ov{P_{\pi}} \right]
\cdot \nn \\
& & \cdot \left( {4 \over \kappa} \psi_m^1 \sigma^{mn} D_n \eta^1
+ i \sqrt{3 \over 2} F^{\hat{5} n} \psi_n^1 \eta^1 \right) +
{\rm \; h.c.}
\, , 
\label{variation}
\eea
where $D_n \eta^1$ contains the spin connection $\omega_{nab}$.
In writing eq.~(\ref{variation}), we exploit the fact 
that $\omega_{na\hat{5}}$ vanishes on the branes, consistently 
with the bosonic jump conditions.

The variation (\ref{variation}) can be cancelled by modifying the
transformation laws of $\psi_5^2$,
\be
\delta \psi_5^2 = \left. \delta \psi_5^2 \right|_{old}
+ 2 \kappa^2 \left[ \delta(x^5) \ov{P_0} + \delta
(x^5 - \pi \kappa) \ov{P_{\pi}} \right] \eta^1 \, , 
\label{modsusy}
\ee
where $\left. \delta \psi_5^2 \right|_{old}$ is as in
eq.~(\ref{susygr}).  With this new transformation, the
bulk action is not invariant,
\bea
\delta S_{bulk} & = & - {\kappa^2 \over 2} \,
\int d^4 x \int_{- \pi \kappa}^{+ \pi \kappa}
d x^5 \, e_4 \left[ \delta(x^5)
\ov{P_0} + \delta(x^5 - \pi \kappa) \ov{P_{\pi}} \right] \cdot
\nn \\ & & \cdot
\left( {4 \over \kappa} \psi_m^1 \sigma^{mn} D_n \eta^1 + i \sqrt{3
\over 2} F^{\hat{5} n} \psi_n^1 \eta^1 \right) 
+ {\rm \; h.c.} \, ,
\eea
where again we exploit the fact that $\omega_{na\hat{5}}$ vanishes at
the fixed points. If we define the total action to be $S = S_{bulk} +
S_{brane}$, the variation $\delta S_{bulk}$ precisely cancels
$\delta S_{brane}$, for any values of $\ov{P_0}$ and $\ov{P_{\pi}}$.

\section{The super-Higgs effect}

In the previous section we constructed a five-dimensional
bulk-plus-brane action that is supersymmetry invariant.  The brane
action is purely fermionic, so the superpotential vevs do not change
the bosonic equations of motion.  In particular, the branes remain
tensionless, so the bosonic background does not warp.

In this section we study the supersymmetry breaking induced by the
superpotential vevs.  We give a complete description of the
super-Higgs effect and the resulting spectrum for all the gravitino
modes.  The initial part of our discussion is close to the treatment
of \cite{hor,mno}, but avoids many of the complications that arise
from Horava-Witten theory, such as the additional moduli of the
Calabi-Yau manifold, the warp factors, etc.

We start by searching for solutions to the Killing spinor equations,
which are determined by the right-hand sides of the supersymmetry
transformations, evaluated in the bosonic background.  The only
nontrivial equations arise from the variations of $\psi_5^1$ and
$\psi_5^2$.  We find
\bea {2 \over \kappa} \partial_5 \eta^1 &=& 0 \, , \nonumber\\ {2
\over \kappa} \partial_5 \eta^2 &=& - 2 \kappa^2 \left[ \delta(x^5)
\ov{P_0} + \delta (x^5 - \pi \kappa) \ov{P_{\pi}} \right] \eta^1 \, .
\eea 
These equations have no solution on the circle except when $\ov{P_0} =
- \ov{P_{\pi}}$.  In this case supersymmetry is preserved, and
\be
\eta^1 = \zeta\ ,\qquad\qquad
\eta^2 = -\kappa^3 \,\Big({\ov{P_0} - \ov{P_{\pi}}\over 4}\Big)\,
\epsilon(x^5)\,\zeta \, ,
\label{kilspi}
\ee
is the Killing spinor for constant $\zeta$; $\epsilon(x^5)$ is the
`sign' function defined on $S^1$.  When $\ov{P_0} \ne - \ov{P_{\pi}}$,
there is no Killing spinor, and supersymmetry is broken spontaneously.
The amount of supersymmetry breaking is fixed by the order parameter
$F = \kappa (\ov{P_0} + \ov{P_{\pi}})$.

This analysis indicates that the supersymmetry breaking is controlled
by the vevs of the superpotentials on the two branes.  These vevs are
determined independently, by the physics on each brane, separated by a
finite distance in the $x^5$ direction.  In this sense the order
parameter for supersymmetry breaking is nonlocal, as in \cite{hor}.

The non-locality of the order parameter makes it worthwhile to study
in detail how the supersymmetry breaking is realized.  In particular,
one would like to identify the Goldstone fermion and investigate the
super-Higgs effect.  To see how this works, we focus on the fermion
bilinears and set all the bosonic fields to their background values.
We find
\bea
S_{2f}^{(5)}
& = &
{1 \over 2} \, \int d^4 x \int_{- \pi \kappa}^{+ \pi \kappa} d x^5
\,
\left\{
{r \over \kappa} \epsilon^{mnpq} \left( \ov{\psi_m^1}
\ov{\sigma}_n \partial_p \psi_q^1 - \psi_m^2 \sigma_n \partial_p
\ov{\psi_q^2} \right) \right.
\nn \\ & & \nn \\ &&
+
{2 \over \kappa} e_4
\left( - \psi_m^2 \sigma^{mn} \partial_5 \psi_n^1
+ \ov{\psi_m^1} \ov{\sigma}^{mn} \partial_5 \ov{\psi_n^2}
+ \psi_m^2 \sigma^{mn} \partial_n \psi_5^1 \right.
\nn \\ & &  \nn \\ & &
\left.
-  \ov{\psi_m^1} \ov{\sigma}^{mn} \partial_n \ov{\psi_5^2}
+ \psi_5^2 \sigma^{mn} \partial_m \psi_n^1
- \ov{\psi_5^1} \ov{\sigma}^{mn} \partial_m \ov{\psi_n^2} \right)
\nn \\ & & \nn \\ &&
+ \left. \left(  e_4 \kappa^2  \left[ \delta(x^5)
\ov{P_0} + \delta(x^5 - \pi \kappa) \ov{P_{\pi}}  \right]
\psi_m^1 \sigma^{mn} \psi_n^1 + {\rm \; h.c.} \right)
\right\} \, .
\label{s52f}
\eea
We then write this action in terms of four dimensional fields, which
we define through a Fourier expansion:
\bea
\psi^+ (x^5) & = & {1 \over \sqrt{\pi r}} \left[ \psi^{+}_{0} +
\sqrt{2} \sum_{\rho=1}^{\infty} \psi^{+}_{\rho} \cos\rho M_5 x^5
\right] \, , \nn \\
\psi^- (x^5)
& = &
{1 \over \sqrt{\pi r}} \left[
\sqrt{2} \sum_{\rho=1}^{\infty} \psi^{-}_{\rho} \sin\rho M_5 x^5
\right] \, .
\label{modez2}
\eea
In this expression, $\psi^+$ stands for $(\psi_m^1,\psi_5^2)$ and
$\psi^-$ for $(\psi_m^2,\psi_5^1)$.  The expansion is consistent with
the boundary conditions and the $Z_2$-parity assignments for the
fields.  We substitute these expressions into (\ref{s52f}) and
integrate over $x^5$ to obtain
\bea
{\cal L}_{2f}^{(4)}
& = & \left\{
{1 \over 2} \epsilon^{mnpq}\left({\ov{\psi}}^1_{p,0}
{\ov{\sigma}}_q \partial_m \psi^1_{n,0}+\sum_{\rho=1}^{\infty}
{\ov{\psi}}^1_{p,\rho} {\ov{\sigma}}_q \partial_m \psi^1_{n,\rho}
+\sum_{\rho=1}^{\infty} {\ov{\psi}}^2_{p,\rho} {\ov{\sigma}}_q
\partial_m \psi^2_{n,\rho} \right) \right. \nn \\
& + &
{2 \over r} e_4 \left( \psi^2_{5,0} \sigma^{mn} \partial_m
\psi^1_{n,0} + \sum_{\rho=1}^{\infty} \psi^2_{5,\rho} \sigma^{mn}
\partial_m \psi^1_{n,\rho} - \sum_{\rho=1}^{\infty} \psi^1_{5,\rho}
\sigma^{mn} \partial_m \psi^2_{n,\rho} \right) \nn \\
& + & {2 \over r} e_4 \sum_{\rho=1}^{\infty} \left( \rho
M_5 \right) \psi^2_{m,\rho} \sigma^{mn} \psi^1_{n,\rho} \nn \\
& + & {\kappa^2 \over 2 \pi r} e_4 \, \ov{P}_0 \left[
\psi^1_{m,0} + \sqrt{2} \sum_{\rho=1}^{\infty}
\psi^1_{m,\rho} \right] \sigma^{mn} \left[
\psi^1_{n,0} + \sqrt{2} \sum_{\sigma=1}^{\infty}
\psi^1_{n,\sigma} \right] \nn  \\
& + & \left. {\kappa^2 \over 2 \pi r} e_4 \, \ov{P}_{\pi}
\left[ \psi^1_{m,0} + \sqrt{2} \sum_{\rho=1}^{\infty}
(-1)^{\rho} \psi^1_{m,\rho} \right] \sigma^{mn} \left[
\psi^1_{n,0} + \sqrt{2} \sum_{\sigma=1}^{\infty}
(-1)^{\sigma} \psi^1_{n,\sigma} \right] \right\} \nn\\
&+&   {\rm \; h.c.} 
\label{l42f}
\eea

This expression shows that the brane superpotentials induce 
mixings between the different Fourier modes.  These mixings
considerably complicate the discussion of the super-Higgs 
effect, as in \cite{mno}. For generic values of $P_0$
and $P_\pi$, with  $P_0 \ne - P_{\pi}$, the fields 
$\psi^1_{5,\rho}$, $\psi^2_{5,0}$ and $\psi^2_{5,\rho}$ 
($\rho>0$) are all goldstinos. They are absorbed by the 
gravitinos through the following field transformations:
\bea
\label{tra1}
\psi^1_{m,0}&\to&\psi^1_{m,0}+\frac{2\pi}{\kappa^2 ({\ov{P}}_0
+{\ov{P}}_\pi)}
\partial_m \psi^2_{5,0} +\sum_{\rho=1}^{\infty}\frac{\sqrt{2}}
{\rho M_5} \left[\frac{{\ov{P}}_0 +(-1)^\rho {\ov{P}}_\pi} {{\ov{P}}_0
+{\ov{P}}_\pi}\right] \partial_m \psi^1_{5,\rho} \, , \\
\label{tra2}
\psi^1_{m,\rho}&\to&\psi^1_{m,\rho}-\frac{1}{\rho M_5} \partial_m
\psi^1_{5,\rho} \, , \\
\psi^2_{m,\rho}&\to&\psi^2_{m,\rho}-\frac{\sqrt{2}}{\rho M_5}
\left[\frac{{\ov{P}}_0+(-1)^\rho {\ov{P}}_\pi}
{{\ov{P}}_0+{\ov{P}}_\pi}\right] \partial_m \psi^2_{5,0}+\nn\\
\label{tra3}
&+&\sum_{\sigma=1}^{\infty} \frac{1}{\pi\rho\sigma}
\frac{{\ov{P}}_0 {\ov{P}}_\pi} {({\ov{P}}_0+{\ov{P}}_\pi)}
\left[1-(-1)^\rho\right]\left[1-(-1)^\sigma\right] \partial_m
\psi^1_{5,\sigma}+\frac{1}{\rho M_5} \partial_m \psi^2_{5,\rho} \, .
\eea
These transformations define the ``unitary gauge'': they
eliminate all the terms containing $\psi^1_{5,\rho}$,
$\psi^2_{5,0}$ and $\psi^2_{5,\rho}$ ($\rho>0$), from
the Lagrangian (\ref{l42f}).  Moreover, they permit us to read the
infinite-dimensional gravitino mass matrix directly
from eq.~(\ref{l42f}).

When $P_0=-P_\pi$, the transformations (\ref{tra1}-\ref{tra3}) are
singular and it is not possible to remove all the modes of $\psi^1_5$
and $\psi^2_5$ from the Lagrangian.  There is one linear combination
that remains massless,
\bea
\psi^{(0)}_n &=& \alpha\left(\psi^1_{n,0}-\sqrt{2}~\frac{\kappa^3}{\pi}
\ov{P_0}\sum_{\rho=0}^{\infty}\frac{1}{2\rho+1}\psi^2_{n,2\rho+1}\right)
\nn \\[2mm]
&=&  \alpha\, M_5 \sqrt{{r\over 4\pi}}
       \int^{+\pi/M_5}_{-\pi/M_5} dx^5\,
       \Bigg[\psi^1_n +\left({\kappa^3 \ov{P_0}\over2}\right)
\epsilon(x^5) \psi^2_n\Bigg]\ ,
\eea
where $\alpha=(1+\kappa^6 |P_0|^2/4)^{-1/2}$ is a normalization.  This
form is consistent with the solution to the Killing spinor equations,
which fixes the gravitino zero mode to be
\be
\psi_n^1 = \psi^{(0)}_n\ ,\qquad\qquad
\psi_n^2 = -\frac{\kappa^3 \,\ov{P_0}}{2}\,
\epsilon(x^5)\,\psi^{(0)}_n\ .
\ee
The massless gravitino indicates that four-dimensional $N=1$
supersymmetry is left unbroken.  There is also one two-component
massless fermion that is not absorbed by the super-Higgs mechanism.
It is described by the combination
\bea
\psi^{(0)}_5 &=& \alpha\left(\psi^2_{5,0}+\sqrt{2}~\frac{\kappa^3}{\pi}
\ov{P_0}\sum_{\rho=0}^{\infty}\frac{1}{2\rho+1}\psi^1_{5,2\rho+1}\right)
\nn \\[2mm]
&=&  \alpha\, M_5 \sqrt{{r\over 4\pi}}
       \int^{+\pi/M_5}_{-\pi/M_5} dx^5\,
       \Bigg[\psi^2_5 -\left({\kappa^3 \ov{P_0}\over2}\right)
\epsilon(x^5) \psi^1_5\Bigg]\ .
\eea
An additional linear transformation in the space ($\psi^{(0)}_n,
\psi^{(0)}_5$) is needed to diagonalize the kinetic terms for the
massless fermions.

In the appendix we derive the gravitino mass spectrum. Taking for
simplicity $P_0$ and $P_\pi$ to be real, we find:
\be
\label{eigvaltxt}
{\cal M}_{3/2}^{(\rho)} = {\rho \over R} + 
{\delta_0 + \delta_\pi \over 2 \pi R} \, , 
\;\;\;\;\; (\rho=0,\pm1,\pm2,\ldots) \, ,
\label{spectrum}
\ee 
where
\be
\delta_{0 \, (\pi)} = 2 \,
\arctan {\kappa^3 P_{0\,(\pi)} \over 2} \, .
\label{deltas}
\ee
Note that when $(P_0 + P_\pi) \ne 0$, the gravitino masses are shifted
with respect to their supersymmetric partners.  Moreover, the lightest
gravitino has a non-vanishing mass.  These facts show that supersymmetry
is indeed spontaneously broken.

\section{Relation to coordinate-dependent compactifications}

Our bulk-plus-brane construction gives a mass spectrum that is
reminiscent of the conventional Scherk-Schwarz mechanism \cite{ss}, in
which all fields are smooth, and the boundary conditions are twisted
by a global symmetry of the five-dimensional theory.\footnote{Some
similarities between gaugino condensation in Horava-Witten theory and
the Scherk-Schwarz mechanism were noticed in \cite{aqgau}.}  The twist
shifts each of the eigenvalues of the gravitino mass matrix by the
same amount, exactly as in (\ref{spectrum}).

The analogy between the two cases is not limited to the fermionic
spectrum.  In each case, the bosonic fields have the same masses
as when supersymmetry is unbroken; the bulk and brane contributions
to the vacuum energy vanish classically; and the compactification
radius is a classical flat direction (together with its super-partner, the
axionic phase associated with the fifth component of the graviphoton).
There is, however, an important difference.  In the conventional
Scherk-Schwarz mechanism, the universal mass shift arises from a
bulk mass term.  In our construction, the shift arises from mass terms
localized at the orbifold fixed points.  The localized masses induce
mixings between all levels of the Kaluza-Klein decomposition.

As we have shown in a companion paper \cite{bfz}, a suitable
generalization of the Scherk-Schwarz mechanism can give rise to
the localized mass terms.  The generalization makes use of twisted
boundary conditions, as in the usual Scherk-Schwarz mechanism,
but allows the fields to have cusps and discontinuities (or `jumps')
at the orbifold fixed points.  In this section we will see that the
five-dimensional supergravity action, with smooth gravitinos and
twisted boundary conditions, is equivalent to a bulk-plus-brane
action, with periodic gravitinos and jumps at the orbifold fixed points.

We start by recalling the essential features of the conventional
Scherk-Schwarz mechanism, for the case of five-dimensional Poincar\'e
supergravity compactified on $S^1/Z_2$.  The Lagrangian has a global
$SU(2)_R$ invariance, under which $\Phi_M \equiv
(\psi^1_M,\psi^2_M)^T$ transforms as a doublet.  [$\Phi_M$ should not
be confused with $\Psi_M \equiv (\psi^1_M, \ov{\psi^2_M})^T$.]  The
gravitino boundary conditions are twisted by a $U(1)_R \subset
SU(2)_R$ transformation,
\be
\Phi^c_M(x^5+2 \pi\kappa)
=
{e^{\dd - i \beta \sigma^2}} 
\Phi^c_M(x^5) \, ,
\label{twist}
\ee
where $\sigma^2$ is a Pauli matrix acting on the space of $(\psi^1_M,
\psi^2_M)^T$.  This twist is consistent with the orbifold projection
defined in eqs.~(\ref{peven}) and (\ref{podd}).  The label `$c$'
indicates that the fields are continuous across the orbifold fixed
points,
\be
\Phi^c_M (+\xi) = \Phi^c_M (- \xi) \, ,
\;\;\;\;\;
\Phi^c_M (\pi \kappa + \xi) =  
\Phi^c_M (\pi \kappa - \xi) \, ,
\;\;\;\;\;
(0 < \xi \ll 1) \, .
\ee

The twisted boundary conditions break the four-dimensional supersymmetry.
To see how this works, we change to gravitino fields
$\widetilde{\Phi}_M(x^5)$ that are periodic on the circle.  The twisted fields
are related to the untwisted fields by
\be
\Phi^c_M(x^5)= V(y) \,
\widetilde{\Phi}_M(x^5) \, ,
\label{ansatz}
\ee
where
\be 
V(y) = \exp\left({\dd - { \dd i \beta \sigma^2 x^5 \over 
\dd 2 \pi \kappa}}\right) \, .  
\ee 
We then substitute (\ref{ansatz}) into the Lagrangian (\ref{lbulk}).  The only
new terms are those in which the $x^5$ derivatives act on the fermionic
fields.  We find
\be 
V^\dagger \partial_5 \Phi^c_M =
\left[ \partial_5 + V^\dagger \partial_5 V \right] \widetilde{\Phi}_M
\equiv \widetilde{D}_5 \widetilde{\Phi}_M \, ,
\ee 
which implies that $\widetilde{D}_5 \widetilde{\Phi}_M $ is a covariant
derivative, with constant connection \cite{hoso}
\be 
A_5 \equiv V^\dagger \partial_5 V = 
- {i \beta \sigma^2 \over 2 \pi \kappa} \, .  
\ee
The connection gives rise to a supersymmetry-breaking gravitino mass
term, one that shifts the gravitino spectrum at each mass level:
\be
{\cal M}_{3/2}^{(\rho)} = {\rho \over R} - {\beta \over 2 \pi R} \, ,
\;\;\;\;\; (\rho=0,\pm1,\pm2,\ldots) \, .  
\ee

We are now ready to show that the bulk-plus-brane action has an alternative
interpretation in terms of a generalized Scherk-Schwarz mechanism  \cite{bfz}.
Because we have compactified on the orbifold $S^1/Z_2$, the gravitino
boundary conditions are characterized by an overall twist {\it and}
by discontinuities at the orbifold fixed points.  We start with the
conventional Scherk-Schwarz fields $\Phi^c_M$, with twist parameter
$\beta$.  We then perform the following field redefinition:
\be
\label{nome}
\Phi^c_M (x^5) = e^{\dd i \alpha(x^5) \sigma^2}
\Phi_M (x^5) \, , \ee
where 
\be
\alpha(x^5) = \frac{\delta_0-\delta_\pi}{4} \epsilon(x^5)
+\frac{\delta_0+\delta_\pi}{4}\eta(x^5) \, .
\ee
In this expression, $\epsilon(x^5)$ is the `sign' function, and
\be
\label{eta}
\eta(x^5) = 2 n + 1 \, , 
\;\;
n \pi \kappa < x^5 < (n+1) \pi\kappa  \, ,
\;\;
(n\in Z) \, ,
\ee
is the `staircase' function that steps by two units every $\pi \kappa$
along $x^5$.

{}From these expressions, it is not hard to check that the fields
$\Phi_M (x^5)$ obey the following jump conditions at the orbifold
fixed points:
\be
\label{jumpcond}
\Phi_M (+\xi) =  e^{\dd i \delta_0 \sigma^2}
\Phi_M (- \xi) \, ,
\;\;\;\;\;
\Phi_M (\pi \kappa + \xi) =  
e^{\dd i \delta_\pi \sigma^2}
\Phi_M (\pi \kappa - \xi) \, .
\ee
The fields $\Phi_M (x^5)$ also have twist $\beta + \delta_0 + \delta_\pi$.  Indeed,
if we choose
\be
\beta = - \left( \delta_0 + \delta_\pi \right) \, ,
\ee
the fields $\Phi_M (x^5)$ are periodic.

The bulk action is not invariant under this field redefinition.  As before, the
$x^5$ derivatives give rise to a connection $A_5$.  Now, however, the
connection is singular; it generates a brane action that is localized at the
orbifold fixed points,
\be
\label{sbrane3}
S_{brane} = {1 \over 2 \kappa} \,
\int d^4 x \int_{- \pi \kappa}^{+ \pi \kappa}
d x^5 \,  e_4 \left[ \delta(x^5)
\delta_0 + \delta(x^5 - \pi \kappa) \delta_\pi  \right]
\, \left( \psi_a^1 \sigma^{ab} \psi_b^1 +
\psi_a^2 \sigma^{ab} \psi_b^2 \right) +
{\rm \; h.c.} 
\ee

Supersymmetry invariance of the total action $S=S_{bulk}+S_{brane}$
is guaranteed by the fact that we have redefined the fields of an invariant
bulk action. The supersymmetry transformations for the fields $\psi_5^{1,2}$
are easily derived,
\bea
\delta \psi_5^1 &=& \left. \delta \psi_5^1 \right|_{old}
- 2 \kappa^2 \left[ \delta(x^5) \ov{P_0} + \delta
(x^5 - \pi \kappa) \ov{P_{\pi}} \right] \eta^2 \, , \nn\\
\delta \psi_5^2 &=& \left. \delta \psi_5^2 \right|_{old}
+ 2 \kappa^2 \left[ \delta(x^5) \ov{P_0} + \delta
(x^5 - \pi \kappa) \ov{P_{\pi}} \right] \eta^1 \, ,
\label{modsusy12}
\eea
where $\left. \delta \psi_5^{1,2} \right|_{old}$ is as in
(\ref{susygr}).  For the special parameter choice $\ov{P_0} = -
\ov{P_{\pi}}$, supersymmetry is not broken, and the Killing
spinor is given by
\be
\eta^1 = \cos \left( {\displaystyle \epsilon(x^5) \kappa^3 \ov{P_0} \over
\displaystyle 2} \right)
\, \zeta \, ,
\qquad\qquad
\eta^2 = - \sin \left( {\displaystyle \epsilon(x^5) \kappa^3 \ov{P_0} 
\over \displaystyle 2} \right) 
\, \zeta \, ,
\label{kilspin}
\ee
in analogy with (\ref{kilspi}).

This discussion exactly parallels the one we gave for the
conventional Scherk-Schwarz mechanism.  However, as
explained in \cite{bfz}, the brane action (\ref{sbrane3}) is
inconvenient for deriving the equations of motion.  The fields
$\psi_m^{1,2}$ are too singular to apply the naive variational
principle without regularization.  Indeed, as explained in
\cite{bfz}, the even fields are not piecewise smooth.  (For
example, $\psi_m^1(0) \ne \lim_{\xi \rightarrow 0} [\psi_m^1(+\xi)
+\psi_m^1(-\xi)]/2$, so one cannot apply the standard Fourier
decomposition.)

Therefore we follow ref.~\cite{bfz} and consider the
equivalent brane action
\be
\label{sbrane4}
S_{brane} = {1 \over \kappa} \,
\int d^4 x \int_{- \pi \kappa}^{+ \pi \kappa}
d x^5 \,  e_4 \left[ \delta(x^5) \tan {\delta_0
\over 2} + \delta(x^5 - \pi \kappa) \tan 
{\delta_\pi \over 2}  \right]
\, \psi_a^1 \sigma^{ab} \psi_b^1 +
{\rm \; h.c.} 
\ee
With this action, the even fields $\psi_m^1$ are continuous,
so we can apply the naive variational principle and derive the
equations of motion.  It is immediate to show, using (\ref{deltas}),
that the brane action (\ref{sbrane4}) coincides with our original
brane action (\ref{sbrane2}).  By integrating the equations
of motion associated with (\ref{sbrane4}), one can derive the
discontinuities of the odd fields,
\be
\label{discbis}
\psi_a^2(+\xi)-\psi_a^2(-\xi)=-2 \tan\frac{\delta_0}{2} \,
\psi_a^1(0) \, ,
\;\;\;\;\;
\psi_a^2(\pi \kappa +\xi)-\psi_a^2(\pi \kappa -\xi)=
-2 \tan\frac{\delta_\pi}{2} \, \psi_a^1 (\pi \kappa)\, ,
\ee
and check that they precisely reproduce the jumps of eq.~(\ref{jumpcond}).

This analysis shows that supersymmetry breaking by brane superpotentials
has an alternative description in terms of generalized boundary conditions
on the gravitino fields.   The supersymmetry breaking is spontaneous because
every gravitino becomes massive via a super-Higgs effect.  The order
parameter, $F \equiv (\delta_0 + \delta_\pi ) / \kappa^2 = - \beta / \kappa^2$,
is manifestly non-local:  in one description, it is related to the Scherk-Schwarz
twist; in the other, it contains contributions from each of the two fixed points.  

As explained in \cite{bfz}, one can further generalize this description by
allowing for a Scherk-Schwarz twist {\it and} for jumps at the orbifold fixed
points. This would give two types of gravitino mass terms: one localized at
$x^5 = (0,\pi \kappa)$, and the other constant in the bulk.

\section{Conclusions and outlook}

In this paper we presented a bulk-plus-brane action that describes
spontaneously broken supersymmetry in $4+1$ dimensions. Supersymmetry
breaking is induced by the expectation values of superpotentials on
tensionless branes. The order parameter for the supersymmetry breaking
is nonlocal; it is determined by the mismatch of the superpotential
vevs on the two branes. The gravitino fields are periodic, but the
equations of motion force the odd fields to be discontinuous at the
locations of the branes. The construction reproduces the main features
of gaugino condensation in $M$-theory at the level of an effective
five-dimensional Lagrangian.

We also showed that our construction is equivalent to a
coordinate-dependent compactification on the orbifold $S^1/Z_2$, where
the gravitino fields and their derivatives are continuous across the
orbifold fixed points but obey twisted boundary conditions.  The
resulting spectrum is identical to that of a conventional
Scherk-Schwarz compactification, for an appropriate choice of the
twist parameter.
 
At low energies and in the limit of small supersymmetry breaking,
the massive Kaluza-Klein modes can be integrated out to give a 
four-dimensional effective action for the light graviton and
radion supermultiplets.  The calculation is relatively easy
because the brane action only affects the fermionic fields.
The bosonic action is a consistent truncation of the one for
the zero modes of five-dimensional supergravity, compactified 
on a circle $S^1$.  It corresponds to a four-dimensional no-scale
supergravity model, with one chiral supermultiplet, whose
interactions are determined by the usual $SU(1,1)/U(1)$
K\"ahler potential \cite{noscale}.
The fermionic terms are fixed by the K\"ahler potential, 
together with a constant superpotential whose value must
be adjusted to match the mass of the lightest gravitino.

As is typical in models with extra dimensions and supersymmetry
breaking, the flat directions are lifted by quantum corrections.
Indeed, it is not hard to compute the one-loop effective potential
for the radion $R$, including contributions from all the Kaluza-Klein
modes.  We find
\bea
V^1_{eff}&=&\frac{1}{2}
\sum_J (-1)^{(2J+1)} {\rm tr} \int\frac{d^4 k}{(2 \pi)^4}
\log (k^2+{\cal M}_J^2)\\
&=&-\frac{1}{8\pi^2} \sum_{n\in Z}
\int_0^{+\infty} \frac{dt}{t^3} \left[e^{-n^2 t/R^2}-
e^{-(n+a)^2 t/R^2}\right]\ ,
\label{eff1}
\eea
where ${\cal M}_J^2$ denotes the squared mass matrix for the particles
of spin $J$.  For the case at hand, the graviton and gravitino masses
are separated by the constant $a/R$, where
\be
a=
{\dd{\delta_0 + \delta_\pi \over 2 \pi}} \, .
\ee
After performing a Poisson resummation we find
\be
V^1_{eff}=-\frac{3}{32 \pi^6 R^4}\left[\zeta(5)-Li_5(e^{2i\pi a})
+ {\rm \; h.c.} \right]\ .
\ee
The result is finite, despite the divergence occurring at
each level in the sum of eq.~(\ref{eff1}).  Note that
the potential has a minimum at vanishing $R$.  In a more
realistic model, a nontrivial minimum at a finite non-zero
value of $R$ can be obtained by adding matter with appropriate 
gauge and Yukawa couplings \cite{kpz}.

Note that the four-dimensional effective theory fails to explain
the special ultraviolet properties of the model, connected
with the non-local character of supersymmetry breaking.  In the
four-dimensional theory, one finds a quadratically divergent
contribution to the one-loop vacuum energy.  The Kaluza-Klein
modes of the five-dimensional theory provide the appropriate
cutoff for the four-dimensional calculation.

The work presented here is a first step towards a more
complete understanding of bulk-plus-brane supersymmetry
breaking.  In this paper, the Goldstone fermions are all
bulk fields.  In a more general scenario, the Goldstone
fermions can involve brane fields as well.  The presence
of $F$- and $D$-terms on the branes might well induce
non-vanishing brane tensions, which would then require
that the bulk background be warped.  Such scenarios are
presently under investigation.

\bigskip

\acknowledgments

We thank J.-P.~Derendinger, K.~Dienes, E.~Kiritsis, C.~Kounnas 
and M.~Porrati for discussions. We also thank the Aspen Center 
of Physics, where part of this work was done, for its warm 
hospitality.  J.B.\ is supported by the U.S. National Science 
Foundation, grant NSF-PHY-9970781. F.F.\ and\ F.Z. acknowledge 
the partial financial support of the European Program 
HPRN-CT-2000-00148.

\bigskip

\bigskip

\appendix

\section{Appendix}

The gravitino mass eigenvalues can be found
by solving the five-dimensional equations of motion,
subject to the boundary conditions specified in section
5, or by diagonalizing the gravitino mass matrix.
In this appendix we take the second approach.

We start with the gravitino mass matrix, which we extract
from (\ref{l42f}).  We then make the following redefinitions:
\be
\psi_{m,\rho}^{\pm} = {\psi_{m,\rho}^1 \pm \psi_{m,\rho}^2
\over \sqrt{2}} \, , \; (\rho > 0) \, ,
\;\;\;\;\;\;
P_{\pm} = {\kappa^3 \over 2 \pi}
\left( \ov{P}_0 \pm \ov{P}_{\pi} \right)
\, .
\ee
This gives
\be
{\cal M}_{3/2} = {1 \over R}
\left(
\begin{array}{c|cc|cc|c}
\cP_+ & \cP_- & \cP_- & \cP_+ & \cP_+ & \ldots \\
\hline
\cP_- & \cP_+ + 1 & \cP_+ & \cP_- & \cP_- & \ldots \\
\cP_- & \cP_+ & \cP_+ - 1 & \cP_-       & \cP_- & \ldots \\
\hline
\cP_+ & \cP_-  & \cP_- & \cP_+ + 2 & \cP_+ & \ldots \\
\cP_+ & \cP_-     & \cP_- & \cP_+ & \cP_+ - 2 & \ldots \\
\hline
\ldots & \ldots & \ldots & \ldots & \ldots & \ldots \\
\end{array}
\right) \,,
\ee
in the basis $(\psi_0^1,\psi_1^+,\psi_1^-,\psi_2^+,\psi_2^-,\ldots)$.

We find the mass eigenvalues by extending the techniques
of ref.~\cite{ddg}.  For simplicity, we take $P_0$ and $P_{\pi}$
to be real. Defining ($E \equiv even$, $O \equiv odd$)
\be
S_E \equiv \sum_{k \in E} ( a_k^+ +  a_k^- ) \, ,
\;\;\;\;\;
S_O \equiv \sum_{k \in O} ( a_k^+ +  a_k^- ) \, ,
\ee
and considering for the moment the dimensionless matrix
$\hat{{\cal M}} \equiv {\cal M}_{3/2} \cdot R$, we rewrite the
eigenvalue equations as
\bea
\cP_+ a_0 + \cP_+ S_E + \cP_- S_O = \lambda a_0 \, ,
&& (n=0) \, , \nn \\
\cP_+ a_0 + \cP_+ S_E + \cP_- S_O = (\lambda + n) a_n^- \, ,
&& (n \in E) \, , \nn \\
\cP_+ a_0 + \cP_+ S_E + \cP_- S_O = (\lambda - n) a_n^+ \, ,
&& (n \in E) \, , \nn \\
\cP_- a_0 + \cP_- S_E + \cP_+ S_O = (\lambda + n) a_n^- \, ,
&& (n \in O) \, , \nn \\
\cP_- a_0 + \cP_- S_E + \cP_+ S_O = (\lambda - n) a_n^+ \, ,
&& (n \in O) \, .
\eea
After a series of manipulations we find:
\bea
\label{uno}
\cP_+ a_0 + \cP_+ S_E + \cP_- S_O = \lambda a_0 \, ,
&& \\
\label{due}
S_E = 2 \lambda ( \cP_+ a_0 + \cP_+ S_E + \cP_- S_O ) \Sigma_E \, ,
&& \\
\label{tre}
S_O = 2 \lambda ( \cP_- a_0 + \cP_- S_E + \cP_+ S_O ) \Sigma_O \, ,
\eea
where:
\be
\Sigma_E \equiv \sum_{n \in E} {1 \over \lambda^2 - n^2}
= - {1 \over 2 \lambda^2} + {\pi \over 4 \lambda} \left[
{ 1 + \cos \left( {\lambda \pi} \right) \over
\sin \left( {\lambda \pi} \right)} \right] \, ,
\ee
\be
\Sigma_O \equiv \sum_{n \in O} {1 \over \lambda^2 - n^2}
= - {\pi \over 4 \lambda} \left[
{ 1 - \cos \left( {\lambda \pi} \right) \over
\sin \left( {\lambda \pi} \right)} \right]
=- {\pi \over 4 \lambda} \tan \left( {\lambda \pi
\over 2} \right) \, .
\ee
Solving (\ref{due}) and (\ref{tre}) for $S_E$ and $S_O$, and
substituting into (\ref{uno}), we find:
\be
4 \pi \cP_+ \cos \left( {\lambda \pi} \right) =
\left[ \pi^2 (\cP_-^2 - \cP_+^2) + 4 \right]
\sin \left( {\lambda \pi} \right) \, ,
\ee
or
\be
\lambda^{(\rho)} = {1 \over \pi} \arctan \left[
4 \pi \cP_+ \over \pi^2 (\cP_-^2 - \cP_+^2) + 4
\right] + \rho  \, ,
\;\;\;\;\;
(\rho=0,\pm1,\pm2,\ldots) \, ,
\ee
Reinstating the overall factor $1/R$, we derive the mass
eigenvalues at each Kaluza-Klein level,
\be
\label{eigval}
{\cal M}_{3/2}^{(\rho)} = {1 \over R}
\left\{ {1 \over \pi}  \arctan \left[
4 \pi \cP_+  \over \pi^2 (\cP_-^2 - \cP_+^2) +
4 \right] + \rho  \right\} \, ,
\;\;\;\;\;
(\rho=0,\pm1,\pm2,\ldots) \, ,
\ee
or, equivalently, eqs.~(\ref{eigvaltxt}) and (\ref{deltas}).
\newpage
\end{document}